\documentclass[journal=acscatal,manuscript=article]{achemso}
\setkeys{acs}{maxauthors = 20}
\usepackage{amsmath}
\usepackage{amstext}
\usepackage{verbatim}

\author{Jingwen Zhou}
\affiliation[ITCC]
{Institute of Theoretical and Computational Chemistry, Key Laboratory of Mesoscopic Chemistry of the Ministry of Education (MOE), School of Chemistry and Chemical Engineering, Jiangsu Key Laboratory of Vehicle Emissions Control, Nanjing University, Nanjing 210023, China}

\author{Yunsong Fu}
\affiliation[ITCC]
{Institute of Theoretical and Computational Chemistry, Key Laboratory of Mesoscopic Chemistry of the Ministry of Education (MOE), School of Chemistry and Chemical Engineering, Jiangsu Key Laboratory of Vehicle Emissions Control, Nanjing University, Nanjing 210023, China}

\author{Ling Liu}
\email{lingliu@pku.edu.cn}
\affiliation[SKLAMMP]
{State Key Laboratory for Artificial Microstructure and Mesoscopic Physics, Frontier Science Center for Nano-optoelectronics and School of Physics, Peking University, Beijing 100871, China}

\author{Chungen Liu}
\email{cgliu@nju.edu.cn}
\affiliation[ITCC]
{Institute of Theoretical and Computational Chemistry, Key Laboratory of Mesoscopic Chemistry of the Ministry of Education (MOE), School of Chemistry and Chemical Engineering, Jiangsu Key Laboratory of Vehicle Emissions Control, Nanjing University, Nanjing 210023, China}

\title[An \textsf{achemso} demo]
{Constant-Potential Machine Learning Molecular Dynamics Simulations Reveal Potential-Regulated Cu Cluster Formation on MoS$_2$}

\keywords{constant-potential quantum chemistry, machine learning force field, single atoms, single cluster, molecular dynamics}

\begin{document}

\begin{tocentry}
	
	\includegraphics[width=8.46 cm,height=4.77 cm]{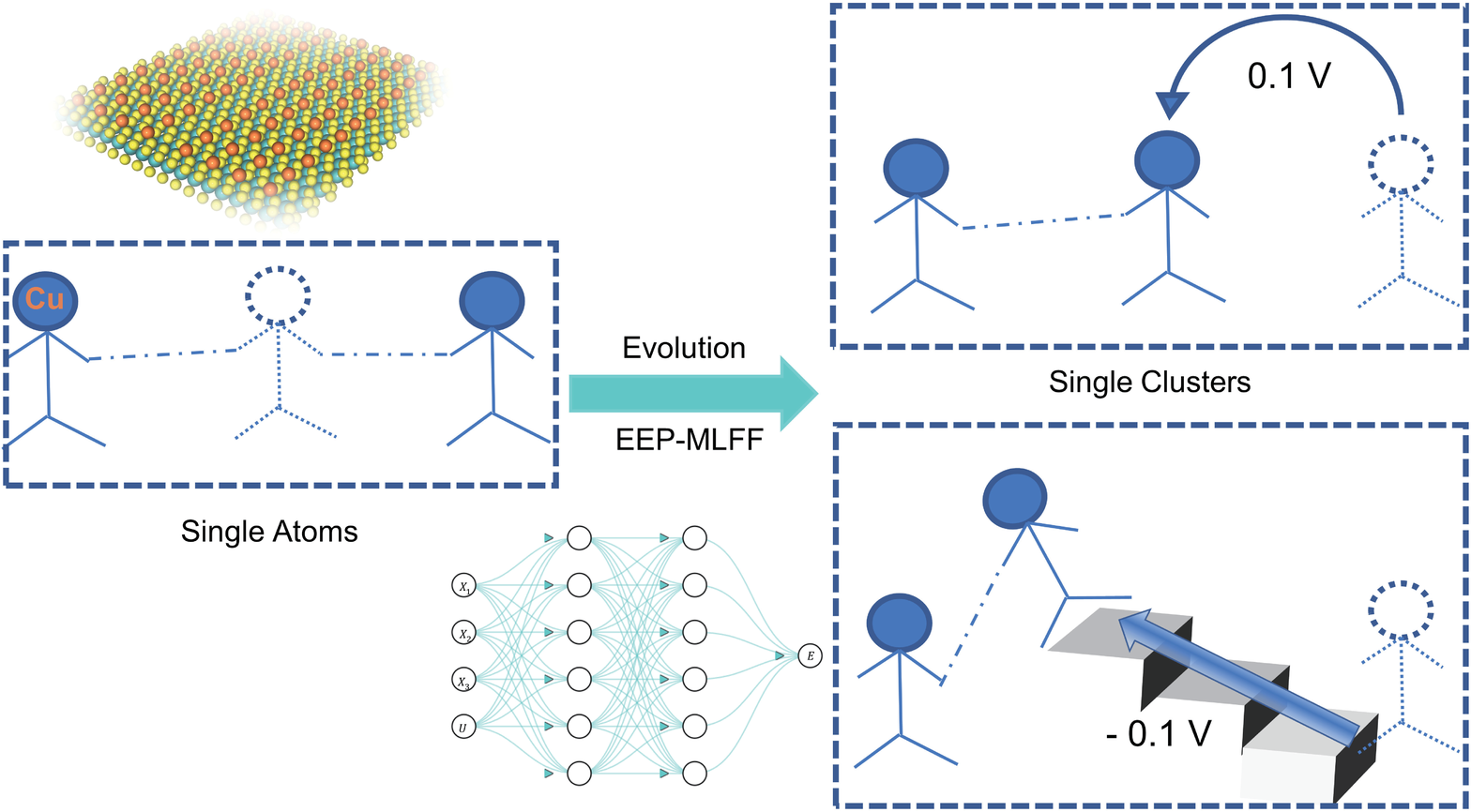}
	
\end{tocentry}




\begin{abstract} 
 
Electrochemical processes play a crucial role in energy storage and conversion systems, yet their computational modeling remains a significant challenge. Accurately incorporating the effects of electric potential has been a central focus in theoretical electrochemistry. Although constant-potential ab initio molecular dynamics (CP-AIMD) has provided valuable insights, it is limited by its substantial computational demands.
Here, we introduce the Explicit Electric Potential Machine Learning Force Field (EEP-MLFF) model. Our model integrates the electric potential as an explicit input parameter along with the atom-centered descriptors in the atomic neural network. This approach enables the evaluation of nuclear forces under arbitrary electric potentials, thus facilitating molecular dynamics simulations at a specific potential. 
By applying the proposed machine learning method to the Cu/1T$^{\prime}$-MoS$_2$ system, molecular dynamics simulations reveal that the potential-modulated Cu atom migration and aggregation lead to the formation of small steric Cu clusters (Single Clusters, SCs) at potentials below -0.1 V. The morphological transformations of adsorbed Cu atoms are elucidated through electronic structure analyses, which demonstrates that both Cu-S and Cu-Cu bonding can be effectively tuned by the applied electric potential. 
Our findings present an opportunity for the convenient manufacture of single metal cluster catalysts through potential modulation. 
Moreover, this theoretical framework facilitates the exploration of potential-regulated processes and helps investigate the mechanisms of electrochemical reactions.
\end{abstract}

\section{Introduction}%

Electrochemistry is a broad field of research that spans from the atomic scale of interfacial reactions to the macroscopic scale of industrial operations.\cite{Yang2022, Yao2022} 
Electrochemical processes are featured as the reactions primarily driven by the applied electric potential bias, which serves as one of the critical thermodynamic variables that significantly influence the kinetics and mechanisms of these reactions. \cite{Bonnet2012}
Therefore, accurately modeling the effect of electric potential is essential for a fundamental understanding of electrochemistry.

In recent years, grand-canonical density functional theory (DFT) methods have been developed and employed to perform molecular dynamics (MD) simulations of constant-potential processes, yielding promising results in accounting for the potential effects. \cite{Bonnet2012, Bouzid2017, Zhao2021, Xia2023a}. 
A key aspect of these methods is the implementation of constant potential condition, which involves using the total number of electrons as an additional degree of freedom that can be optimized to maintain the potential at a preset value. The dynamics of nuclear coordinates and electron numbers are propagated separately, decoupling the optimization of electron numbers from electronic structure calculations.

Although DFT methods have been demonstrated to be effective for gaining atomic-level insights into various processes, simulating large systems beyond hundreds of atoms remains prohibitively expensive, especially for processes occurring on nanosecond timescales.\cite{Scandolo2019}
Consequently, developing highly efficient computational methods, in virtue of the machine learning potentials (MLP), has attracted significant attention to enhance the feasibility of ab initio molecular dynamics (AIMD) simulations.\cite{Bartok2010, Behler2015, Unke2021a}
However, most existing MLP models were not designed to deal with electrochemical systems. The integration of machine learning techniques with constant-potential methods and their application to large-scale MD simulations of electrochemical systems remains largely unexplored. 

In the context of implicit solvent models, it has been well established that the energy of an electrochemical system with varying electric potential can be effectively represented by a quadratic function over a relatively broad range of potentials. \cite{Mathew2019, Liu2018, Liu2020a, Duan2020,Goff2021,Hagopian2022,Liu2020b,Weitzner2017a} %
Consequently, it is natural to incorporate the electric potential into the input layer of neural networks used in machine learning force field models. The strategy is implemented and leads to the development of an Explicit Electric Potential Machine Learning Force Field (EEP-MLFF) model. Our newly proposed model enables constant-potential MD simulations with machine learning potentials, facilitating the large-scale simulations of electrochemical systems. 

After years of major concern on single atom catalysts (SACs),\cite{Qiao2011, Wang2018c} single cluster catalysts (SCCs), have emerged as a significant topic of interest, \cite{Ji2017a, Jia2022, Ma2018b, Liu2020b} albeit the precise synthesis of specially designed SCCs is quite challenging. \cite{Li2023a, Peng2021} 
Recent in-situ and \textit{operando} experimental studies have shown that electric potential can effectively manipulate the dynamic transformations between SACs and SCCs, revealing that the actual active catalytic sites are predominantly small clusters, rather than individual mono-atoms in some cases.\cite{Wei2023,Yang2022a,Zhang2023a} 
In this work, we employ the proposed EEP-MLFF model to simulate the single-atom Cu/1T$^{\prime}$- MoS$_2$ (SA-Cu/1T$^{\prime}$- MoS$_2$) system.
The constant-potential MD simulations uncover the potential-regulated morphological transformation from single-atom Cu (SA-Cu) to single-cluster Cu (SC-Cu). Our findings aim to offer theoretical insights in guiding the experimental synthesis of SCCs.
 
\section{Methodology } 
	
	\begin{figure}
	\includegraphics[width=4.5in]{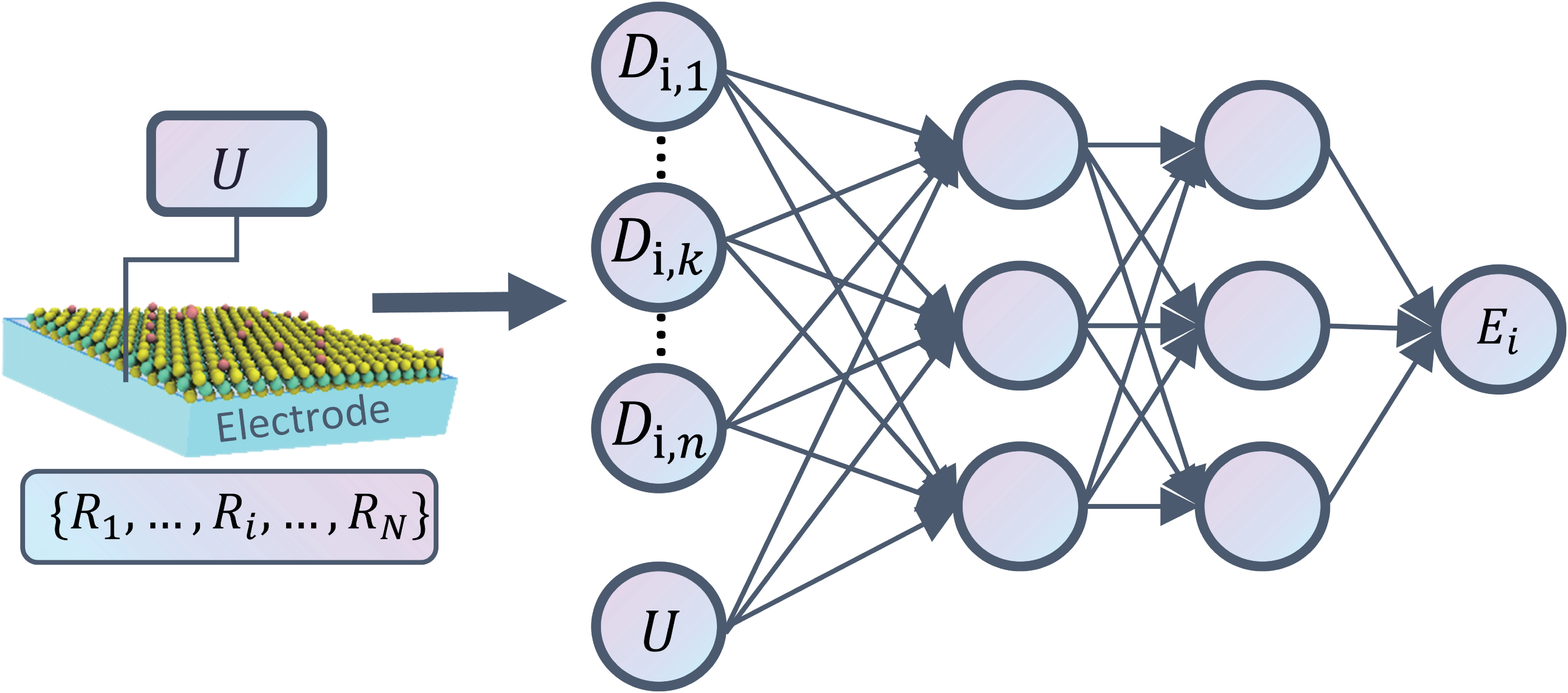}
	\caption{The neural network architecture of the EEP-MLP model. The electric potential $U$ is integrated with traditional atom-centered descriptors to form the input layer.}
		\label{NN}
	\end{figure}
	
	AIMD has primarily been utilized for picosecond-timescale simulations using moderate-sized slab models, due to its high computational cost. 
	In contrast, large-scale simulations of electrochemical systhave recently seen the development of empirical many-body expansion methods, specifically, potential-dependent cluster expansion (CE) models, enabling constant-potential Monte Carlo (MC) simulations.\cite{Tan2013a, Weitzner2017a, Zhou2023} In traditional CE frameworks, the energy of a given system is decomposed into contributions from a series of deliberately designed effective cluster interactions (ECIs). However, in potential-dependent CE models, the potential effect is approximately incorporated, drawing upon the established quadratic relationship between energy and electric potential. Like conventional CE models, the potential-dependent CE models are unable to explicitly handle nuclear coordinates, which precludes the derivation of atomic forces crucial for molecular dynamics simulations. This limitation excludes their use in simulating metal atom aggregation on substrates, where precise tracking of potential-driven geometrical changes is essential.
	
	The highly popular atomic neural network models have been shown to be exceptionally efficient in generating ab initio-level force field models for specific systems through machine learning.
	To enhance the application of machine learning for potential-dependent energies, we introduce the electric potential $U$ as an additional descriptor alongside conventional structural descriptors, forming the EEP-MLFF model. 
	The modification to conventional MLP is illustrated in Figure~\ref{NN}. The total energy is expressed as 	
	\begin{equation}
		E_{\text{tot}}=\sum_{i}^{N}E_i (\mathbf {D}_{i}(\mathbf{R}_1,...,\mathbf{R}_N), U).
		\label{eq_energy}
	\end{equation}	
	where $\mathbf D_i = \{D_{i,k}\}$, denotes the conventional structural descriptors, and the index $k$ labels the descriptor of atom $i$. It is important to note that all atoms are subjected to the same electric potential, ensuring that the value $U$ remains constant for any atomic index $i$. 
	By utilizing the atom-centered descriptors, the potential-dependent nuclear force on atom $i$, $\mathbf{F}_i(\mathbf{R}_1,...,\mathbf{R}_N; U)$, is computed through the partial derivatives of the electric potential-related energy with respect to its nuclear coordinate vector $\mathbf{R}_i$, as shown in the following equation, 
	 \begin{equation}
	 	\mathbf{F}_i(\mathbf{R}_1,...,\mathbf{R}_N; U) = -\frac{\partial E_{\text{tot}}}{\partial \mathbf{R}_i} = -\sum_{j=1}^{N} \frac{\partial E_j(\mathbf{D}_j, U)} {\partial \mathbf{R}_i} = -\sum_{j=1}^{N} \sum_{k=1}^{n} \frac{\partial D_{j,k}}{\partial \mathbf{R}_i} \frac{\partial E_j}{\partial D_{j,k}}. 
	 \label{eq_force}	
	 \end{equation}
Obviously, the EEP-MLFF model can predict the energies and nuclear forces of configurations under specific potentials.
In principle, the selection of structural descriptors is quite flexible. Here, we choose the smooth SO(3) Power Spectrum components to illustrate the construction of the EEP-MLFF model.\cite{Bartok2013} Prior research has shown that employing the smooth SO(3) spectrum descriptor facilitates the development of MLPs capable of accurately capturing the diverse atomic environments encountered in heterogeneous and nanoparticle systems, further simplifying the structural complexity of the neural network model.\cite{Yanxon2021}

In contrast to many constant-potential ab initio molecular dynamics methods,\cite{Bonnet2012, Bouzid2017, Zhao2021, Xia2023a} our approach circumvents the computational overhead of electric potential calibration by explicitly integrating $U$ into the neural network's input layer. Furthermore, the constant-potential treatment is established on the presumption that system charge fluctuations can be fully suppressed within each time step of MD simulation, eliminating significant fluctuation of the electric potentials.\cite{Goldsmith2021} 
Due to the concise expressions of electric potential effects in equations \ref{eq_energy} and \ref{eq_force}, integrating our EEP-MLFF model into various MD simulation packages requires minimal code modifications. In this study, we implement the machine learning model with {PyXtal}\_{FF} package \cite{Yanxon2021} and modify its interface with LAMMPS package,\cite{Thompson2022} to conduct constant-potential machine-learning molecular dynamics (CP-MLMD) simulations.

\section{Results and discussions}

\subsection{Validation of the Machine-Learning Force Field} 

An EEP-MLFF model will be built to explore the potential-driven migration and aggregation of Cu atoms adsorbed on single-layer 1T$^\prime$-MoS$_2$, within an electric potential window from +0.36 V to -0.3 V vs. SHE. 
We employ a slightly wider voltage range of -0.5 V to +0.6 V to train the force field model.
The structures in the training and test sets are generated from AIMD simulations conducted on slab models with varying numbers of adsorbed Cu atoms on the 1T$^\prime$-MoS$_2$ surface. Different amounts of charge are introduced into the systems to simulate varying electric potential conditions. The solvation environment and the overall charge neutrality of the system are achieved using the implicit solvation model VASPsol. \cite{Mathew2014, Mathew2019}
Metadynamics simulations are also employed to generate trajectories that capture the migration of Cu atoms, with the coordination number of Cu atoms serving as the collective variable. All the technical details of simulations can be found in the supplementary information.

The structural dataset comprises a total of 2,999 neutral structures with varying numbers of surface copper atoms (Cu$_1$ to Cu$_12$, Cu$_14$ and Cu$_18$), 900 Cu$_4$/MoS$_2$ slab models with +0.5$e$ charge, and 420 Cu$_5$/MoS$_2$ with -0.5$e$ charge. 
Detailed information on the compositions of these structure sets, alongside their corresponding electric potentials, is provided in Figures S1-S3, Tables S1 and S2. 
\begin{figure}
	\includegraphics[width=0.9\textwidth]{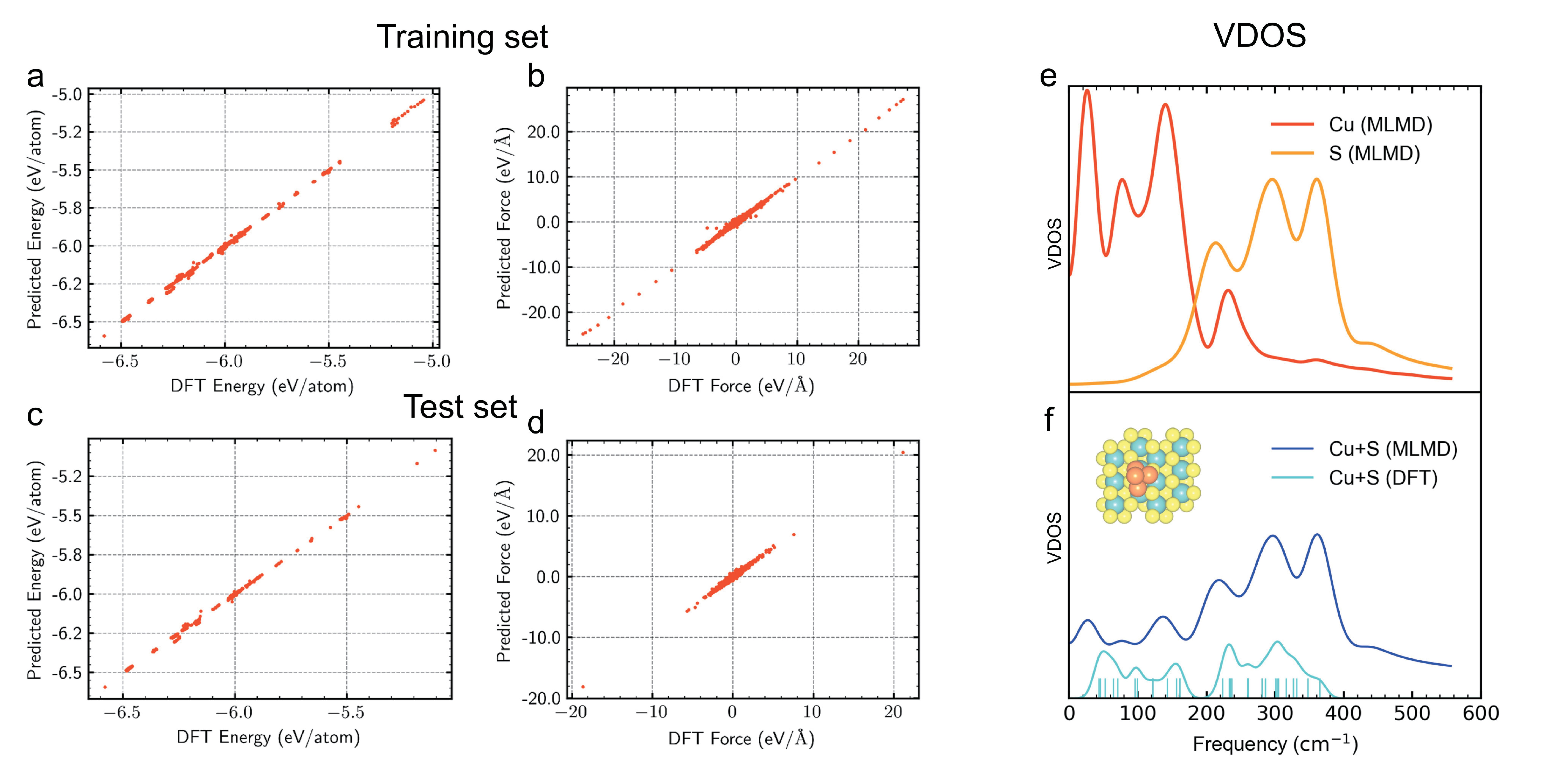}
	\caption{Validation of the EEP-MLFF model in reproducing the results of DFT method. (a)-(d) The accuracy of the EEP-MLFF model in reproducing electronic energy and nuclear forces for structures in the training and test sets. (e)-(f) The signatures of the vibrational spectrum of Cu$_4$/MoS$_2$ evaluated with the constant-potential molecular dynamics simulation at 0 V vs SHE. in (e), the red and yellow lines are the spectra of normalized vibrational density of states (VDOS) of Cu atoms, and their close-contacting S atoms, respectively. in (f), the blue line denotes the simulated total VDOS spectrum of the Cu$_4$ cluster together with the close-contacting S atoms. The cyan vertical lines are vibrational frequencies of these concerned atoms calculated with the DFT method. Each vibrational frequency is broadened using a Gaussian function ($\sigma = 10$) to produce a continuous spectrum (cyan curves). The structure in the inset of Figure~\ref{fig2_data}f displays the DFT-optimized geometry of the Cu$_4$/1T$^\prime$-MoS$_2$ model at 0 V vs. SHE.} 
	\label{fig2_data}
\end{figure}
Figure~\ref{fig2_data}a-d graphically illustrates the accuracy of the EEP-MLFF model in predicting both energy and nuclear forces. In test set, the root mean square errors (RMSE) values of energy and forces are 8 meV/atom, 0.08 eV/$\mathrm \AA$,
demonstrating a good coincidence between the force field and the DFT data. A detailed list of the coefficient of determinations ($R^2$), mean absolute errors (MAE), and RMSE can be found in Table S3.
Since the calculated vibrational frequencies are sensitive to the accuracy of nuclear forces, we further compare the vibrational frequency distribution between DFT and CP-MLMD, with a Cu$_4$ cluster on 1T$^\prime$-MoS$_2$ under an electric potential of 0 V vs. SHE. 
We focus on the vibrational motions of the four Cu atoms as well as the directly contacting five S atoms. The normalized vibrational density of states (VDOS), $g(\nu)$, is obtained by taking the Fourier transform of the velocity autocorrelation functions (VACF) of these atoms, which are then averaged across selected $N$ atoms, \cite{Meyer2011}
	\begin{equation} 
		g(\nu) = \int_{-\infty}^{\infty} dt \frac{\sum_{i=1}^{N} \langle \mathbf{v}_i(t) \cdot \mathbf{v}_i(0) \rangle}{\sum_{i=1}^{N} \langle \mathbf{v}_i(0) \cdot \mathbf{v}_i(0) \rangle} e^{i2\pi \nu t},
	\label{eq_vdos}
	\end{equation}
where $\nu$ denotes the vibrational frequency, and $\mathbf{v}_i(t)$ is the velocity of atom $i$ at time $t$. The simulated VDOS is based on a 60 ps of CP-MLMD simulation trajectory, which has been equilibrated for over 10 ps.
It can be seen from Figure~\ref{fig2_data}e that Cu contributes to the low-frequency vibrational motions, while S contributes to the high-frequency ones. By inspecting Figure~\ref{fig2_data}f, the combined VDOS of Cu and S atoms closely matches the Gaussian-broadened results obtained from the normal mode analysis using the DFT method.

The validity of the EEP-MLFF model is further supported by a CP-MLMD simulation of a much larger model system as shown in the inset of Figure S4, which corresponds to an approximately 7\% coverage of Cu on MoS$_2$. By setting the electric potential to $U=+0.685$V, which is close to the potential of zero charge (PZC) for SA-Cu/MoS$_2$,\cite{Li2022} the isolated metal adsorbates are found to remain well dispersed, avoiding aggregation into clusters. This observation is in accordance with the reported extended X-ray absorption fine structure (EXAFS) experiments.\cite{Li2022} %

\subsection{Evolution From Single Atoms to Single Clusters} 

	\begin{figure}
		\includegraphics[width=6.6 in]{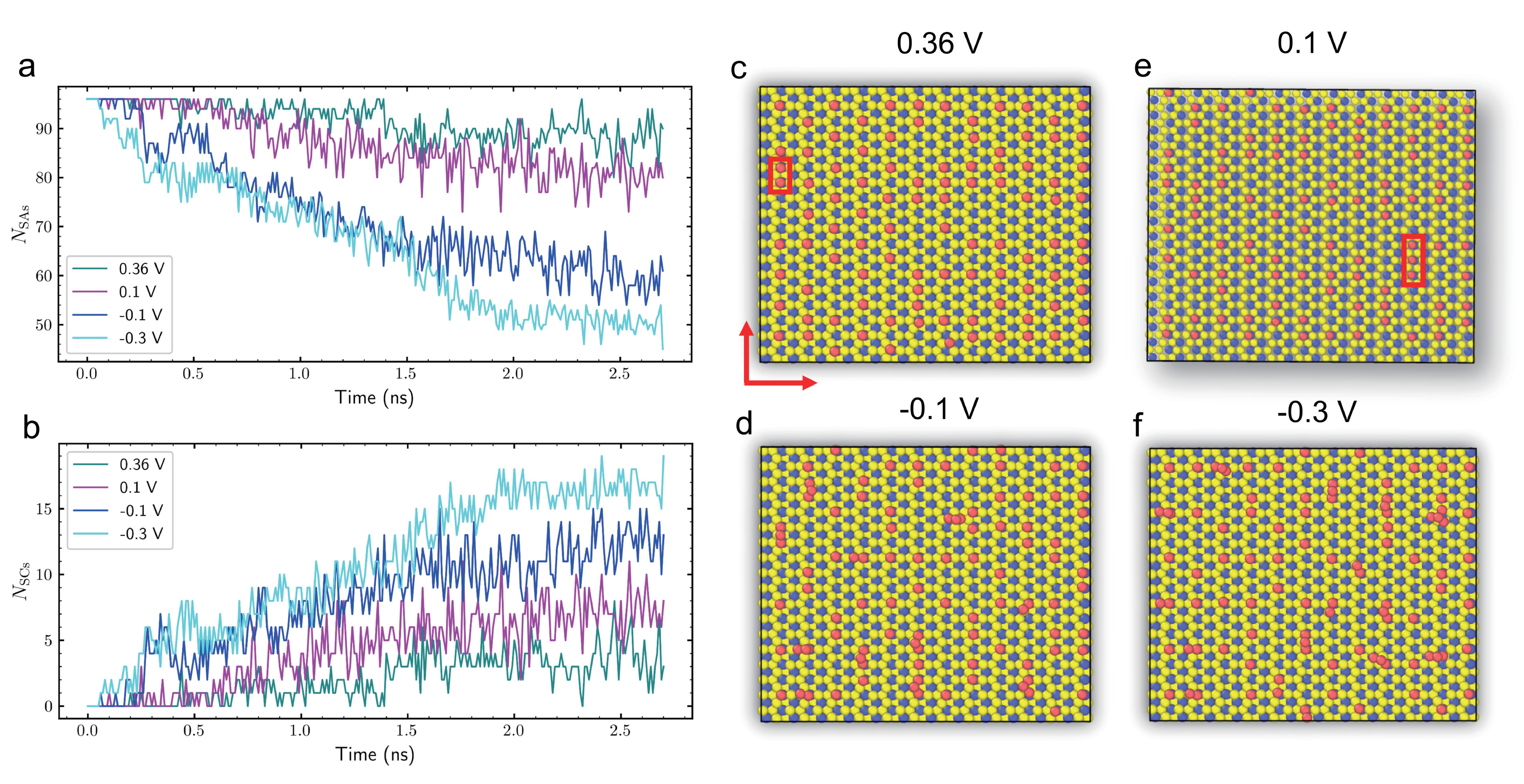}
		\caption{Illustration depicting the morphological evolution of SA-Cu/MoS$_2$ under various electrochemical conditions: (a) The diminishing count of SAs, $N_{\rm {SAs}}$, across molecular dynamics trajectories at differing electric potentials; (b) The emergence and growth of single cluster structures, denoted by $N_{\rm {SCs}}$, within molecular dynamics trajectories at varying electric potentials. (c-f) The snapshots from the MD trajectories simulated with varying electric potentials. The orange, yellow and blue colored balls represent the elements Cu, S and Mo, respectively.}
		\label{fig2_N_CL}
	\end{figure}
		
To elucidate the impact of electric potential on the aggregation of Cu atoms into clusters, CP-MLMD simulations have been performed at varying electric potentials on a large-scale Cu/MoS$_2$ model. The slab model is initialized with 96 independently adsorbed Cu atoms, resulting in approximately 22\% surface coverage of SA-Cu (Figure S5).
It can be seen from Figure~\ref{fig2_N_CL}c-f that snapshots reveal distinct patterns of Cu atom distribution on MoS$_2$ surface, strongly depending on the applied potential. 
As the potential decreases, Cu atoms begin to aggregate, forming short-range ordered atomic arrays (Cu$_2$ and Cu$_3$ with red squares),\cite{Shan2021, Chang2024a} and ultimately develop into small steric clusters when the potential falls below -0.1 V.

To quantify the structural evolution, the formation of clusters is tracked using an interatomic Cu-Cu distance threshold of 3.2 $\rm \AA$, which is analogous to the nearest Mo-Mo distance in the regular 1T$^\prime$-MoS$_2$ structure. A Cu atom is considered as an isolated SA-Cu if its distances with all the other Cu atoms exceed this threshold; otherwise, it forms an SC-Cu together with the other Cu atoms within the distance threshold. 
As shown in Figure S6, employing a more strict Cu-Cu distance threshold of 2.4 \text{\AA} fails to efficiently identify flat-oriented Cu atom array as a single cluster. Therefore, the threshold of 3.2 \text{\AA} is more reasonable.
Figure~\ref{fig2_N_CL} compares the time evolution of the numbers of SA-Cu and SC-Cu species across varying electric potentials. Obviously, the generation of SC-Cu moiety is promoted by the negative potential bias. While approximately 90\% Cu atoms exist as SA-Cu species at 0.36 V, this percentage drops sharply down to less than half when the applied potential is -0.3 V.

Figure S7 presents a more detailed analysis of the competition among different sizes of SC-Cu clusters. 
Surprisingly, the ratios of various clusters do not exhibit a monotonous trend of evolution with increasing negative bias of the electric potential. 
Specifically, under more positive potentials, Cu$_2$ clusters tend to prevail over larger clusters above +0.1 V. However, as the potential decreases to -0.1 V, the proportion of Cu$_3$ becomes comparable to Cu$_2$, accompanied by a non-negligible presence of Cu$_4$ species. Furthermore, as the potential drops to -0.3 V, the ratios of Cu$_3$ and Cu$_2$ continue to rise compared to -0.1 V, while the proportion of larger clusters has essentially disappeared. 
The intriguing observation could suggest a mechanism switch along with the change in electric potential, shift from a thermodynamically controlled slow process at more positive potentials to a kinetically controlled rapid process at more negative potentials. This is further supported by the much faster migration of copper atoms at potentials of -0.1 V and -0.3 V compared to 0.36 V and 0.1 V, through the mean square distance (MSD) analysis (Figure S8).
Accordingly, it is reasonable to suppose that more negative potentials favor the formation of smaller clusters, such as Cu$_2$ and Cu$_3$, potentially lead to a significant depletion of the surrounding SA-Cu atoms and ultimately hindering the further growth of clusters in size.

\subsection{Spatial Configurations of Aggregated Cu Atoms} 

\begin{figure} 
	\includegraphics[width=6in]{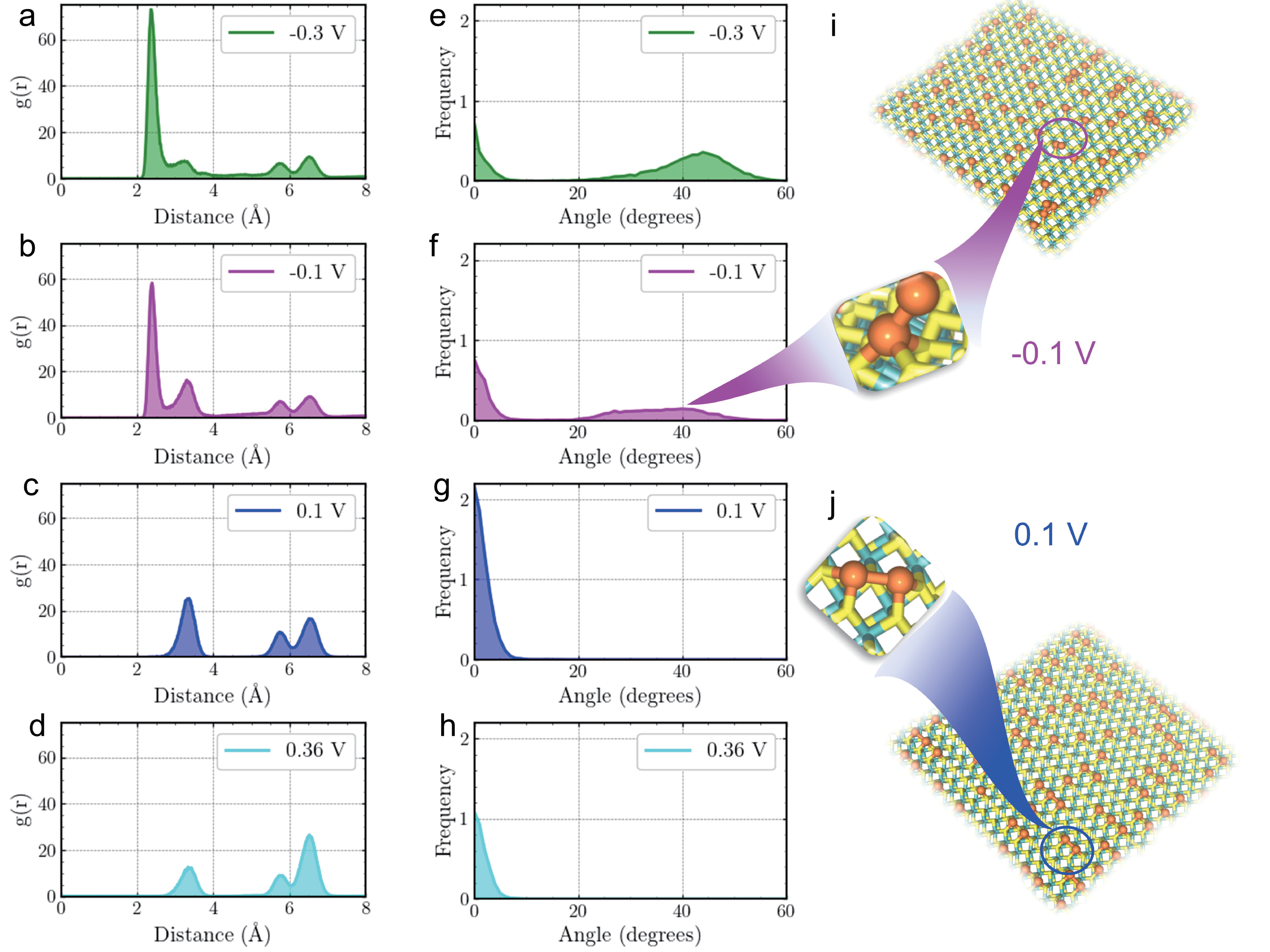}
	\caption{The influence of electric potential on the morphology of adsorbed Cu atoms on MoS$_2$. Panels (a)-(d) depict the radial distribution function of the adsorbed Cu atoms at different potentials, while panels (e)-(h) present the angular distribution function, highlighting potential-driven evolution of the incline angle of the formed Cu dimers relative to the MoS$_2$ surface. Panels (i) and (j) offer magnified views of flat and tilted Cu-Cu dimer configurations, respectively, formed at higher and lower electric potentials.} 
	\label{fig4_RDF}
\end{figure}

Besides the statistical analysis on the time evolution of the formed atomic clusters and arrays, the interatomic radial distribution function (RDF) and the angular distribution function (ADF) offer additional valuable tools for characterizing the structures in a dynamical system. These analyses are based on the last 0.2 ns of CP-MLMD trajectories obtained under varying electric potentials. The RDF at 0.36 V exhibited three characteristic peaks, located at approximately 3.2 \AA, 5.8 \AA, and 6.4 \AA~(Figure~\ref{fig4_RDF}d). 
With increasing distances, these peaks correspond to the Cu-Cu pairs occupying the nearest neighboring adsorption sites in $y$ and $x$ directions, as well as the next nearest neighboring sites along $y$-direction, respectively.
As illustrated in Figure~\ref{fig4_RDF}a-d, when the potential shifts from 0.36 V to 0.1 V, the centers of the three peaks remain essentially unchanged, except for the intensified first peak, indicating an increase in the number of short-range ordered atomic arrays with a growing negative bias of potential. Conversely, when the electric potential is decreased to -0.1 V, a new peak emerges around 2.4 \AA, accompanied by the weakening of the other peaks, suggesting the formation of close-contacting atomic clusters. From -0.1 V to -0.3 V, this new peak continues to intensify, indicating an increase in the number of formed Cu clusters.

Angular distribution analysis is necessary to reveal the formation and evolution of spatial configurations of Cu clusters at different electric potentials. Since Cu$_2$ and Cu$_3$ clusters will dominate at negative potentials, we focus on the distributions of the incline angle of Cu$_2$ from the surface (Figure~\ref{fig4_RDF}e-h) and the angle between the plane where Cu$_3$ is located and the surface plane (Figure S9).
The ADFs of Cu$_2$ clusters present a significant shift in dimer orientation as the electric potential shifts from positive to negative biases.
Specifically, under positive biases, the angles of inclination are tightly clustered around $0^\circ$, indicating a predominantly parallel alignment with the surface of the substrate. When negative biases are applied, a wide range of inclination angles from $20^\circ$ to $60^\circ$ appears, indicating the existence of sterically aggregated Cu$_2$ atoms. The geometries of these moieties are illustrated in Figure~\ref{fig4_RDF}i and g, emphasizing the distinction between conditionally formed tilted and horizontal copper pairs depending on the potential bias. A similar relation between the spatial configurations and potentials is also confirmed in the ADFs of Cu$_3$ cluster.

By integrating the results of simulated MSD in Figure S8 with the aforementioned RDF and ADF analyses, it is reasonable to conclude that the copper atom migration mechanism likely switches at a turning point potential situated between +0.1 V and -0.1 V.
Higher than this turning point potential, the migration of copper atoms is primarily confined to stochastic transitions between neighboring adsorption sites. The attraction among Cu atoms lacks the force necessary to extract adjacent copper atoms from their adsorption sites and bring them together to create sterically aggregated clusters. Nevertheless, the inter-metallic attraction could still facilitate the formation of loosely bound, short-range ordered atomic arrays.
In contrast, when the electric potential falls below this critical threshold, the inter-metallic interaction intensifies, supplying copper pairs in neighboring adsorption sites with an augmented attractive force, which can be sufficiently strong to overcome the Cu-S binding, thus facilitating the creation of sterically aggregated clusters. 

The significance of single cluster moiety in optimizing catalytic performance has been noticed recently, as they were reported to be the primary active catalytic centers in various SAC-catalyzed electrochemical reactions. \cite{Wei2023,Yang2022a,Zhang2023a} Very recently, it has been found that the atom arrays can also exhibit unique catalytic behaviors compared to isolated SACs.\cite{Chang2024a} 
Our simulations indicate that electric potential could influence the aggregation of the SAs, promoting the formation of atom arrays or even steric atomic clusters. This suggests the potential for developing a feasible technical scheme of synthesizing high-performance catalysts by controlling the applied electrical potential, and also reminds us of the importance of \textit{operando} characterizations in exploring the mechanism of electrocatalysis. 

\subsection{Exploring the Potential Induced Electronic Effect in Manipulating Metal Aggregations}

\begin{figure}
	\includegraphics[width=0.7\textwidth]{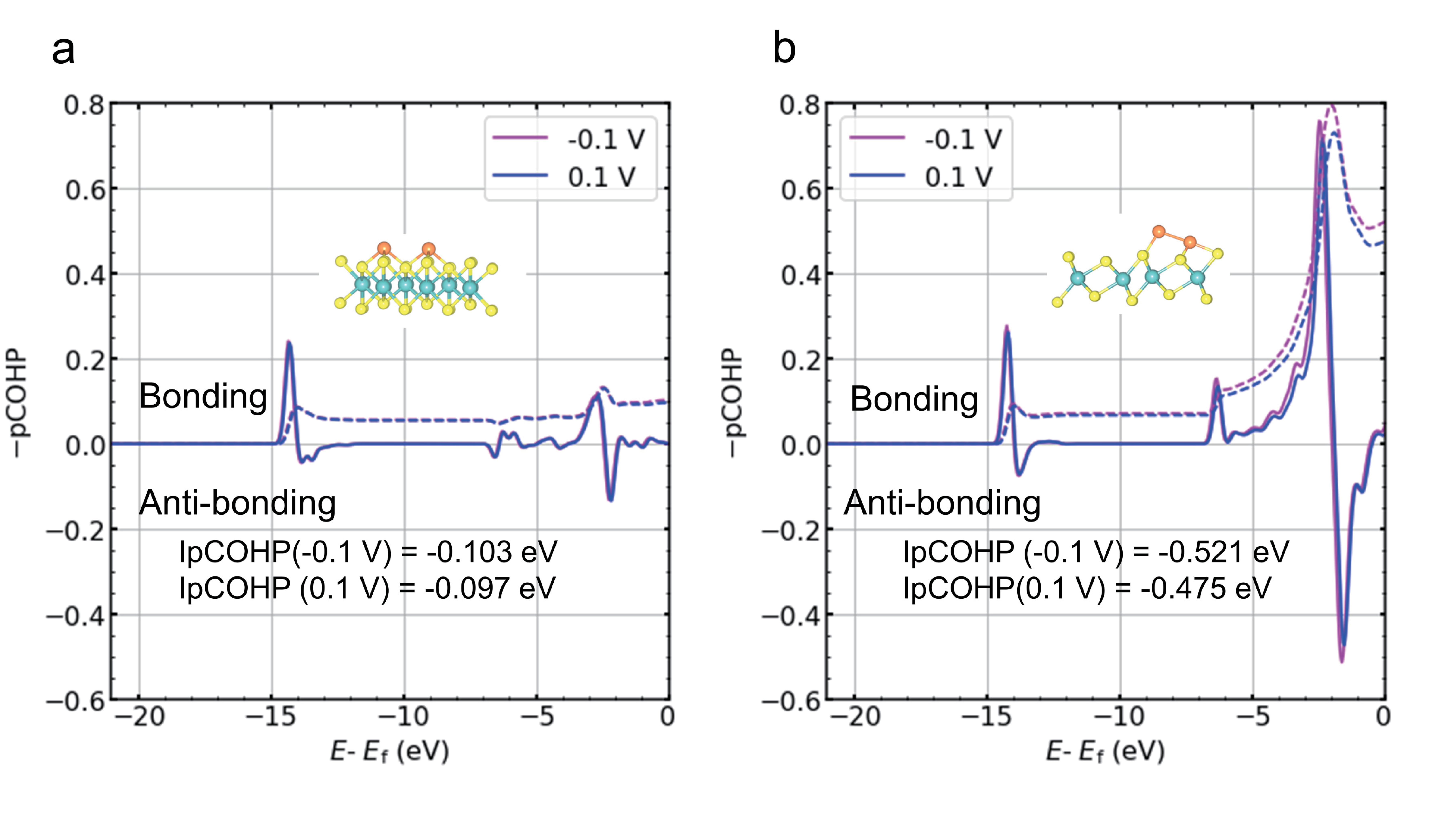}
	\caption{Characterization of the potential effect on the electronic structures of Cu dimers on MoS$_2$ in both flat (a) and tilted (b) configurations through projected crystal orbital Hamilton population (pCOHP) analyses. pCOHP functions (solid lines) and their integration curves (dashed lines) illustrating the effects of morphology and potential on Cu-Cu chemical bonding, which are computed at -0.1 V and 0.1 V. Orange, yellow, and teal spheres represent Cu, S, and Mo elements, respectively.} 
	\label{fig5_es} 
\end{figure}

To explore the intrinsic driving force behind the formation of atomic clusters, we conduct a projected crystal orbital Hamilton population (pCOHP) analysis to quantify the influence of electric potentials on the strengths of chemical bonding of the SA-Cu and the Cu dimer with the MoS$_2$ surface, respectively. \cite{Dronskowski1993, Deringer2011, Maintz2013a}
Figure S10 illustrates the pCOHP results for three Cu-S atom pairs in the SA-Cu system at +0.36 V and -0.3 V. 
Although the electric potential exhibits a minor effect on the pCOHP function across the entire energy range below the Fermi level, discernible distinctions are still observed in the d-band regions, primarily situated above -0.5 eV. Interestingly, the consistently shortened lengths of all three Cu-S bonds at the more negative potential suggest a strengthening of these bonds. However, the slight downward shift of the boundary between the bonding and anti-bonding regions indicates a weakening of the bonds due to an expanded anti-bonding energy region. These counteracting effects lead to a weakening of the two stronger Cu-S bonds while strengthening the weaker one, as reflected in the integrated pCOHP (IpCOHP) shown in Figure S10.
It should be mentioned that this strengthened Cu-S bond does not play a determining role in the migration of the Cu atom on MoS$_2$, as it remains the weakest among all three Cu-S bonds.

For Cu dimer adsorption, pCOHP analyses are performed for both flat and tilted dual-Cu adsorption models at +0.1 V and -0.1 V. 
Figure~\ref{fig5_es}a and b clearly show distinctly different Cu-Cu chemical bonding in the two adsorption configurations. The plots of pCOHP functions as well as their integration curves reveal a significant contribution from electrons in the d-band region near Fermi level in the tilted configuration, which can be attributed to the close contact of the atom pair therein. 
As the potential decreases from +0.1 V to -0.1 V, the integration curves indicate that the negative potential efficiently intensifies the Cu-Cu bonding in the tilted configuration, while having a minor impact on the flat one.

Consequently, applying a more negative electric potential will weaken the stronger Cu-S bonds, thus possibly facilitating the migration of Cu atoms on the MoS$_2$ surface. At the same time, additional charge analysis (Figure S11) reveals that reducing the electric potential will result in an increase in the electron density of these Cu atoms, the oxidation state of the catalytic centers can also affect catalytic activity. 
Besides the above-mentioned morphological effect, it is also noteworthy that the regulation of the electron density on adsorbed atoms accompanying the morphological evolutions could potentially alter the catalytic performance of the active centers. \cite{Mostafa2010a, Mistry2014, Ahmadi2016a,Jin2022b}

\section{Conclusion} 
In conclusion, the electric potential is integrated with conventional atomic-centered descriptors to construct a neural network-based explicit electric-potential machine-learning force field model, EEP-MLFF. 
This straightforward approach to handling potential effects incurs negligible additional computational costs when performing constant-potential molecular dynamics simulations, compared to traditional canonical simulations.
The accuracy of the EEP-MLFF is validated by the satisfactory coincidences in the structure and vibrational spectrum with the DFT method. 
We present an illustrative application of this model, focusing on the simulation of morphological transformations of SA-Cu/MoS$_2$ under electrochemical conditions.
Constant-potential molecular dynamics simulations uncover the potential-modulated Cu atom migration and aggregation, resulting in the formation of steric SC-Cu at potentials below -0.1 V vs SHE, a picture not observed at +0.1 V or more positive potentials. 
According to the chemical bonding analyses in virtue of the projected crystal orbital Hamilton population method, the morphological transformations of adsorbed Cu atoms at negative potentials are elucidated as the consequence of the slight weakening of the Cu-S bonding between the adsorbate and substrate, alongside the strengthening of bonding between closely contacting Cu-Cu pairs.
Our findings suggest the potential for developing a feasible method of manufacturing single cluster catalysts. 
This theoretical framework provides a useful approach for studying potential-regulated processes and offers insights into fundamental electrochemical processes, including electrocatalytic reactions and interfacial particle transport in liquid and solid electrolyte secondary batteries.

\begin{suppinfo}


The Supporting Information is available free of charge.
\begin{itemize}
	\item SI.pdf: Introduction to the theoretical formulation of implementing constant potential molecular dynamics. 
	 Computation details of DFT calculations, the structure and parameters of the neural network for the machine-learning force field, as well as the parameters for implementing MD simulations.
	 Supplementary Figures and Tables, illustrating the construction and constitution of the training and test sets for machine learning. 
	 The radial distribution function of Cu-Cu at the potential of zero charge, which is derived with 0.5 ns of the constant-potential MLMD trajectory after the equilibration period. 
	 The original SA-Cu configuration of Cu/MoS$_2$.
	 The time evolution of the populations of different-size clusters which are computed at different electric potentials.
	 The plots of mean square displacement (MSD) of Cu atoms. 
	 The average dihedral angle distribution function over the last 0.2 ns between the formed Cu trimer plane and the surface (xy plane) under different potentials.
	 The pCOHP analyses between the Cu atom and its three nearest S atoms computed at -0.3 V and 0.36 V.
	 The charge populations for different configurations at 0.1 V and -0.1 V.
	 
	\end{itemize}

\end{suppinfo}

\begin{acknowledgement}
This work was supported by the National Key Research and Development Programs of China (2023YFA1506902, 2022YFA1503103) and the National Natural Science Foundation of China NSF (Grand No. 22073041). We thank the High Performance Computing Center of Nanjing University for computational resources. 
	
\end{acknowledgement}

\bibliography{ref}
	
\end{document}